\documentclass[epj,referee]{svjour}
\usepackage{amsmath,amssymb,graphicx}

\begin{document}
\title{Opinion Dynamics in an Open Community}
\author{Timoteo Carletti\inst{1}, Duccio Fanelli\inst{2,4}, Alessio
  Guarino\inst{3}, Franco Bagnoli\inst{4} \and Andrea Guazzini\inst{4}
}                     
%
%
\institute{1. D\'epartement de Math\'ematique, Facult\'es Universitaires Notre
  Dame de la 
  Paix, 8 rempart de la vierge B5000 Namur, Belgium \email{timoteo.carletti@fundp.ac.be}\\
2. School of Physics and Astronomy, University of Manchester, M13 9PL,
    Manchester, United Kingdom \email{duccio.fanelli@manchester.ac.uk}\\
3. Universit\'e de la Polyn\'esie Francaise, BP 6570 Faa'a, 98702, French
  Polynesia \email{alessio.guarino@upf.pf}\\
4. Dipartimento di Energetica and CSDC, Universit\`a di
  Firenze, and INFN, via S. Marta, 3, 50139 Firenze, Italy
  \email{franco.bagnoli@unifi.it} \email{andrea.guazzini@unifi.it}}
\date{Received: \today}
%
\abstract{We here discuss the process of opinion formation in an open
community where agents are made to interact and
consequently update 
their beliefs. New actors (birth) are assumed to replace individuals
that abandon the community (deaths). This dynamics
is  
simulated in the framework of a simplified model that accounts for mutual
affinity between agents. A rich phenomenology is presented and
discussed  
with reference to the original (closed group) setting. Numerical findings are
supported by analytical calculations. 
} 
\authorrunning{Timoteo Carletti et al.}
\maketitle

\section{Introduction}

Opinion dynamics modeling represents a challenging field where ideas from statistical physics and non-linear science 
can be possibly applied to understand the emergence of collective behaviors,
like consensus or polarization in social groups. Several toy models 
have been proposed in the past to reproduce the key elements that supposedly drive the process of opinion making \cite{opinion} . This large production  
has not been always accompanied by an adequate benchmarking effort to the relevant psychological literature, and doubts are consequently 
being cast on the interpretative ability of the proposed mathematical formulations. Despite these intrinsic limitations, several 
models of interacting agents display a rich and intriguing dynamics which
deserves to be fully unraveled.

Opinion dynamics models can be classified in two large groups. On the one hand, opinions are represented as discrete (spin--like)
variables where the system behaves similarly to spin glasses
models~\cite{Sznajd}. On the other, each individual bears a continuous
opinion   
which span a pre--assigned range~\cite{Deffuant}. In both approaches, a closed system is generally assumed, meaning that the same pool of actors is made to interact during the 
evolution. This can be interpreted by 
  assuming that the inherent dynamical timescales (e.g. opinion convergence
time) are 
much faster than those associated to the processes (e.g. migration,
birth/death) responsible for a modification of the group
composition, 
these latter effects having being therefore so far neglected. Such an implicit assumption is certainly correct when the debate is bound to a small community of individuals, thus making it possible 
to eventually achieve a rapid convergence towards the final configuration. Conversely, it might prove inaccurate when applied to a large ensemble of interacting 
agents, as the process becomes considerably slower and external perturbations need to be accounted for. Given the above, it is therefore of interest 
to elucidate the open system setting, where the population is periodically renewed. 

To this end, we refer to the model presented in~\cite{Bagnoli_prl}, where the role of affinity among individuals is introduced as an additional ingredient. 
This novel quantity measures the degree of inter--personal intimacy and sharing, an effect of paramount importance in real social system~\cite{Nowak} . Indeed, the outcome of an 
hypothetic binary interaction relies on the difference of opinions, previously postulated, but also on the quality of the mutual relationships. 
The affinity is dynamically coupled to the opinion, and, in this respect, it
introduces a memory bias into the system: affinity between agents increases when their opinion tends to converge. 

The aforementioned model is here modified to accommodate for a
death/born like process. In this formulation, $M$ agents
are randomly eliminated from the system, every $T$ time steps. When an
agent exits from the community (virtually, dies), he is immediately replaced by a new element,  
whose opinion and affinity with respect to the group are randomly assigned. As we shall see, the perturbation here prescribed alters dramatically the behavior of the system, with reference to 
the ideal close--system configuration. To understand such modifications via combined numerical and analytical tools, constitutes the object of  
the investigations here reported.


The paper is structured as follows. We first introduce the model, then we present the obtained analytical and numerical results and, finally, we sum up and draw our conclusions.


\section{The model}
\label{sect:model}

In the following, we will shortly review the model previously introduced in ~\cite{Bagnoli_prl} and present the additional features
that are here under inspection. The interested reader can thus
refer to the original paper \cite{Bagnoli_prl}  for a more detailed account on the model characteristics.

Consider a population of $N$ agents and assume that at time $t$ they bear a  scalar opinion $O_i^{t} \in [0,1]$. We also introduce the $N \times N$ time 
dependent matrix ${\bf \alpha}^{t}$, whose elements $\alpha_{ij}^{t}$ belong to the interval $[0,1]$. The quantities $\alpha_{ij}^{t}$  specify the affinity 
of individual $i$ vs. $j$, at time $t$: Larger values of  $\alpha_{ij}^{t}$ are associated to more trustable relationships. 
    
Both the affinity matrix and the agents opinions are
randomly initialized 
time $t=0$.  At each time step $t$, two agents, say $i$ and $j$, are selected
according to the following extraction rule: first the agent $i$ is randomly selected, with a
uniform probability. Then, the agent $j$ which is closer to $i$ in term of the social metric 
$D_{ij}^\eta$  is selected  for interaction. 
The quantity $D_{ij}^\eta$ results from the linear superposition of the so--called social distance,  $d_{ij}$,  
and a stochastic contribution $\eta_j$, namely:
\begin{eqnarray}
 D_{ij}^\eta = d_{ij}^t + \eta_j(0,\sigma)\, .
\label{social_metrics}
\end{eqnarray}

Here $\eta_j(0,\sigma)$ represents a 
normally distributed noise, 
with mean zero and variance $\sigma$, the latter being named social temperature.  
The social distance is instead defined as:
\begin{eqnarray}
d_{ij}^t &=& \Delta O_{ij}^{t} (1-\alpha_{ij}^{t}) \qquad j=1,...,N \qquad j
\ne i\, ,
\label{social_distance}
\end{eqnarray}
with $\Delta O_{ij}^{t}=|O_i^t-O_j^t|$.

The smaller the value of $d_{ij}^t$ the closer the agent $j$ to $i$,
both in term of affinity and opinion. The additive noise $\eta_j(0,\sigma)$  
acts therefore on a fictitious 1D manifold, 
which is introduced to define the pseudo--particle (agent) interaction on the
basis of a nearest neighbors  
selection mechanism and, in this respect, set the degree of mixing in the
community.

When the two agents $i$ and $j$ are selected on the basis of the recipe
prescribed above, they interact and update their characteristics according to
the following scheme~\footnote{The evolution of the quantities $O_j(t)$ 
  and $\alpha_{ij}(t)$ 
is straightforwardly obtained by switching the labels $i$ and $j$ in
the equations.}:  
\begin{equation}
\label{opinion}
  \begin{cases}
O_i^{t+1} &= O_i^{t}- \frac{1}{2} \Delta O_{ij}^{t}
\Gamma_1\left(\alpha^t_{ij}\right) \\ 
\alpha_{ij}^{t+1} &= \alpha_{ij}^{t} + \alpha_{ij}^{t}
      [1-\alpha_{ij}^{t}] \Gamma_2 \left(\Delta O_{ij}\right) \, ,
  \end{cases}
\end{equation}
where the functions $\Gamma_1$ and $\Gamma_2$ respectively read:
\begin{equation}
  \label{eq:Gamma1}
  \Gamma_1 \left(\alpha^t_{ij}\right)= \frac{1}{2}\left[ \Theta (
  \alpha_{ij}^{t}-\alpha_c) + 1 \right]
\end{equation}
and
\begin{equation}
  \label{eq:Gamma2}
  \Gamma_2 \left(\Delta O_{ij}\right)= -\Theta(|\Delta O_{ij}^{t}| - \Delta
O_c)
\end{equation}
and the symbol $\Theta(\cdot)$ stands for the Heaviside
step--function~\footnote{In   
\cite{Bagnoli_prl}, the switchers $\Gamma_1$ and $\Gamma_2$ are smooth
functions constructed from  
the hyperbolic tangent. We shall here limit the discussion to considering the
Heaviside approximation, which is  
recovered by formally sending $\beta_{1,2}$ to infinity in Eqs. (3) and (4)
of~\cite{Bagnoli_prl}.}. More specifically,    
$\Gamma_{1}$ is $0$ or $1$ while $\Gamma_{2}$ is $-1$ or 
$1$, depending on the value of their respective arguments.  In the above
expressions $\alpha_c$ and $\Delta
O_c$ are constant 
  parameters. Notice that, for $\alpha_c
  \rightarrow 0$, the opinion and affinity are formally decoupled
  in~\eqref{opinion}, and the former evolves following the Deffuant et 
  al. scheme~\cite{Deffuant} with $\mu=0.5$. 

\begin{figure*}
\centering
\includegraphics[width=7cm]{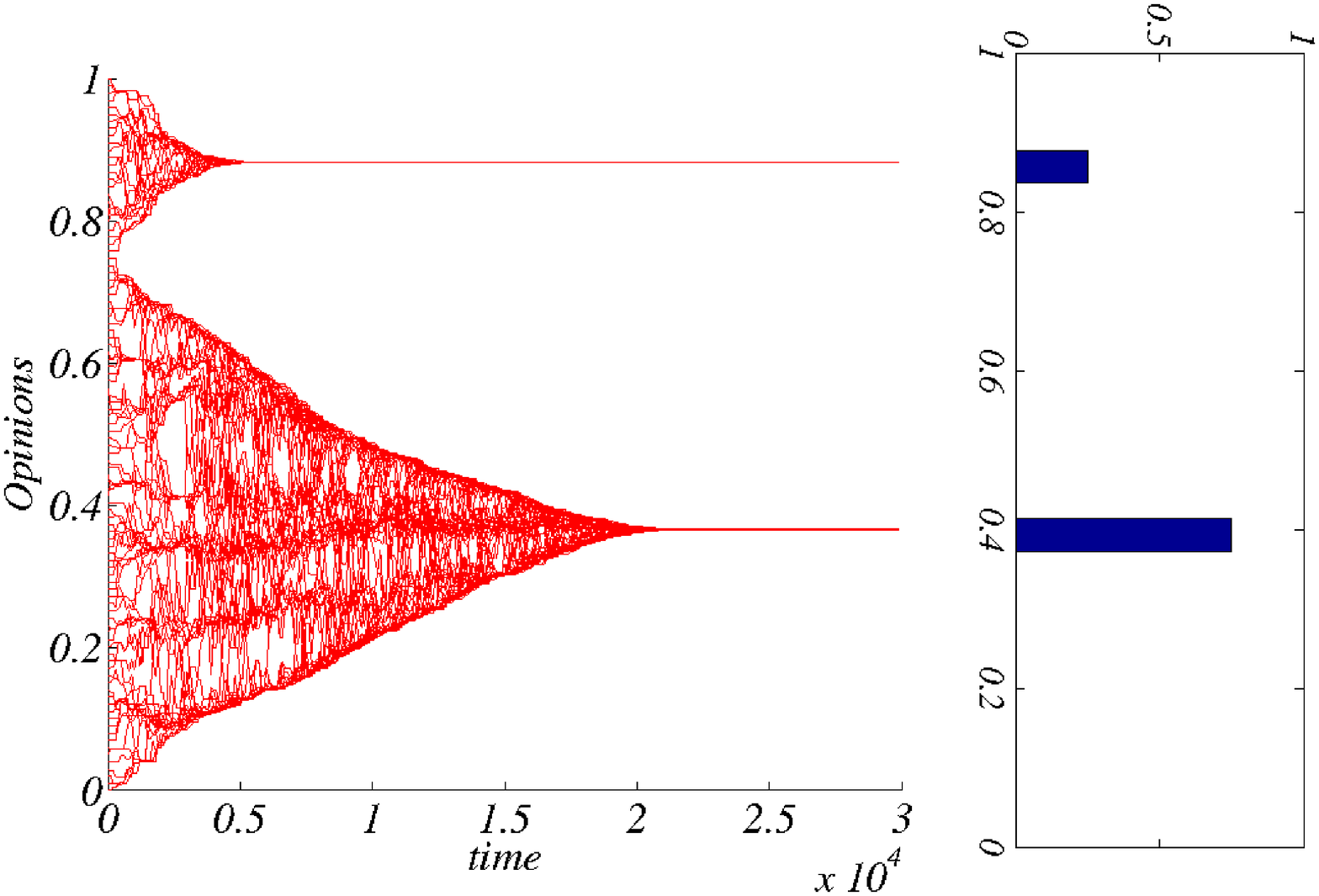}
\includegraphics[width=7cm]{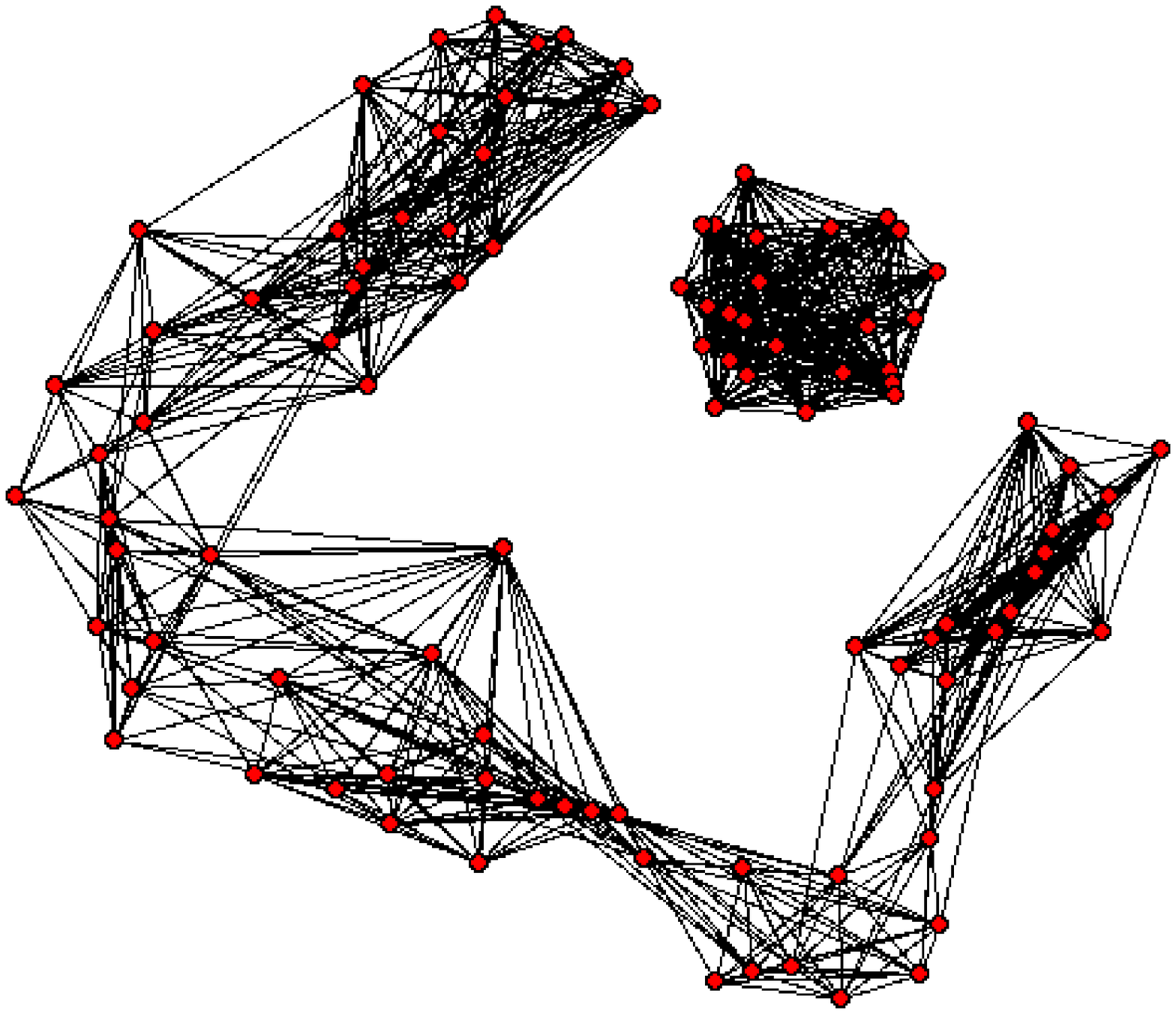}
\caption{Left panel: Typical evolution of the opinion
  versus time, 
i.e. number of iterations and the asymptotic distribution of agents' opinion
in a closed community. Right panel: Final
affinity matrix, represented by the underlying network: node $i$ is linked to
node $j$ if $\alpha_{ij} > \bar{\alpha}=0.8$. Here 
$\sigma= 4 \cdot 10^{-4}$, $\Delta O_c = 0.5$, $\alpha_c = 0.5$ and
$\rho=0$, i.e. no agent can leave the group. Initial opinion are
(random) uniformly distributed within the interval $[0, 1]$, while
$\alpha_{ij}^0$ 
is initialized with uniform (random) values between $0$ an $0.5$.}
\label{fig_qualitativa}
\end{figure*}

In ~\cite{Bagnoli_prl}, a complete analysis of the qualitative behavior of the model as a function of the involved parameters is reported. Asymptotic clusters of opinion are formed, each agglomeration being different in size and centered
around distinct opinion values. Individuals sharing the same believes are also 
characterized by a large affinity scores, as it is exemplified in
Figure~\ref{fig_qualitativa}.  

More quantitatively, the system is shown to undergo a continuous phase transition: above a critical value of the control parameter
$(\sigma \alpha_c)^{-1/2}$ the system fragments into several opinion clusters, otherwise convergence to a single group is numerically 
shown to occur. The interested reader can refer to \cite{Bagnoli_prl} for a speculative interpretation of this finding in term of its psychological
implications. We shall here simply notice that a significant degree of mixing (large social temperature $\sigma$) brings the system towards the single--cluster
final configuration.   

Starting from this setting, we introduce the birth/death process, which in turn amounts to place the system in contact with an external reservoir. The perturbation
here hypothesized is periodic and leaves the total number of agent unchanged. Every $T$ time steps (i.e. encounter events) $M$ agents, randomly selected, are forced to
abandon the system (death). Every removed individual is instantaneously replaced by a new element, whose initial opinion and affinity are randomly fished, with
uniform probability, from the respective intervals $[0,1]$ and $[0, \alpha_{max}]$. Further, we introduce $\rho=\frac{M}{T}$ to characterize the departure density, a crucial
quantity that will play the role of the control parameter in our subsequent developments. As a final remark, it should be emphasized that no aging mechanisms 
are introduced: agents are mature enough to experience peer to peer encounters from the time they enter the system.

\section{Results}
\label{sect:result}

Numerical simulations are performed and the evolution of the system monitored. Qualitatively, the system shows the typical critical behavior and asymptotic 
clusters of opinion, as observed in its original formulation \cite{Bagnoli_prl}. However, peculiar distinctions are found, some of those being addressed in the
following discussion. 
 
\begin{figure}[htbp]
\centering
\includegraphics[width=10cm]{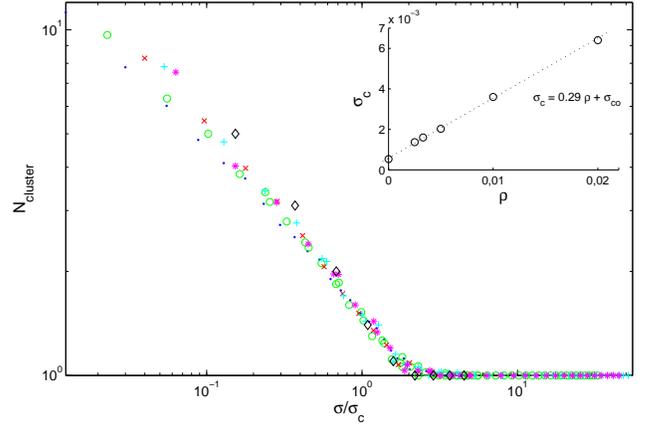}
\caption{Main plot: Average number of clusters as function of the rescaled
  quantity $\frac{\sigma}{\sigma_c}$ for different values of the density   
$\rho$. Diamonds, stars, crosses, x-marks, circles
and solid diamonds represent, respectively, simulations run at $\rho=0$,  
$\rho=0.0025$, $\rho=0.0033$, $\rho=0.005$, $\rho=0.01$ and $\rho=0.05$. In
all simulations, here and after reported, unless otherwise specified,   
$\alpha_c = \Delta O_c = 0 .5$; $O_i^0$ and $\alpha_{ij}^0$ are random
variables uniformly distributed in the intervals $[0,1]$ and  
$[0,\alpha_{max}]$ -- being $\alpha_{max}=0.5$ -- respectively. Inset : $\sigma_c$ as a function of $\rho$. The open circles represent the values calculated from the transitions shown in the main plot. The dotted lines represent the best linear fit : $\sigma_c=0.3 \rho + \sigma_{co}$, with $\sigma_{co}=5.5 \cdot 10^{-4}$. Notice that $\sigma_c=\sigma_{co}$ is eventually recovered in the closed-system setting, which 
in turn corresponds to $\rho=0$.} 
\label{fig1}
\end{figure}

First, an apparently smooth transition is also observed within this novel
formulation, which divides the mono-- and multi--clustered
phases. Interestingly,  
the transition point is now sensitive to the departure density $\rho$. To
further elucidate this point, we draw in Figure~\ref{fig1} the average number
of  
observed clusters versus a rescaled temperature. A clear transition towards
an ordered (single-clustered) phase is observed,  
as the temperature increases. The parameter $\sigma_c$ in Figure~\ref{fig1} plays the role of
an effective temperature, and it is numerically 
adjusted to make distinct curves collapse onto the same profile, which hence
holds irrespectively of the value of $\rho$. The inset of Figure~\ref{fig1}
shows that there is a 
linear correlation between $\sigma_c$ and $\rho$. The larger the departure density $\rho$, the larger the effective temperature $\sigma_c$. In other words, 
when $\rho$ is made to increase (i.e. the system is experiencing the effect of a more pronounced external perturbation), one needs to augment the degree of mixing,
here controlled by the social temperature $\sigma$, if a convergence to the final mono--cluster is sought.  
The death/birth process here modeled is in fact acting
against the thermal contribution, which  brings into contact otherwise  
socially distant individuals. While this latter effect enhances the chances of
convergence, the former favors the opposite tendency to spread.

\begin{figure*}
\centering
\includegraphics[width=15cm]{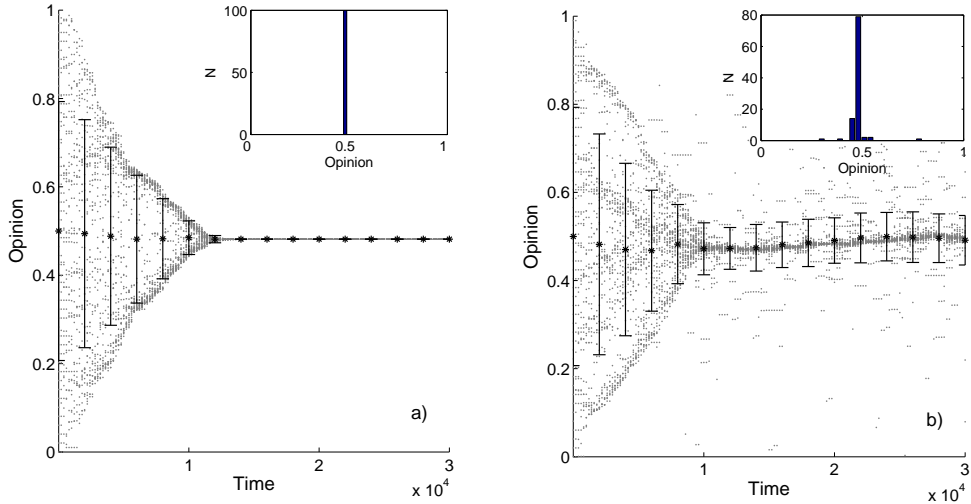}
\caption{Opinion evolution versus
  time, the latter being quantified through the number of iterations. The black stars represent the mean opinion and the error bar the standard deviation of the opinion distribution. In the insets, the histogram of asymptotic
  distribution of agents' opinion.  
In panel a) the system is closed (i.e. $\rho=0$), in b) $\rho=5 \cdot
10^{-3}$.}  
\label{fig2}
\end{figure*}

To further elucidate the role of the external perturbation, we shall refer to
the dynamical regime where the agents converge to a  single  
cluster. When $\rho$ is set to zero, the final shared opinion is 0.5 to which
all agents eventually agree, see Figure~\ref{fig2}a. In other words, the final 
distribution is a Dirac delta, with the peak positioned in
$O=0.5$. Conversely, for $\rho>0$, the final distribution of opinions presents
a clear spreading around  
the most probable value, still found to be $0.5$. This scenario is clearly
depicted in Figure~\ref{fig2}b.  

\begin{figure}[htbp]
\centering
\includegraphics[width=9cm]{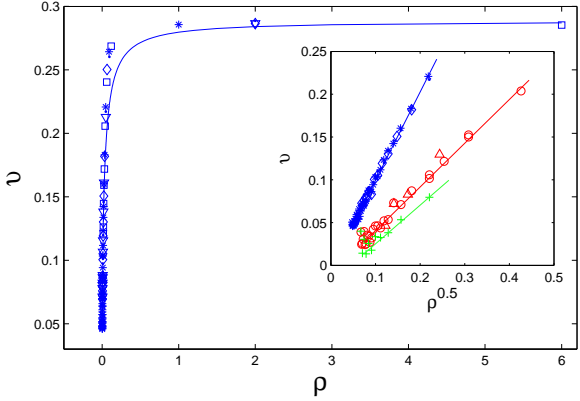}
\caption{Main: The
  standard deviation of the final mono--cluster $\upsilon$ as 
  a function of the departure density $\rho$. Each point results from
  averaging out $100$  
independent runs. Symbols refer to numerical simulation performed with
different values of $\alpha_c$, $M$ and $\sigma$ (stars: $\alpha_c=0.5, M=1,
\sigma=0.07$, diamonds: $\alpha_c=0.5$, $M=2$, $\sigma=0.07$, squares:
$\alpha_c=0.5$, $M=6$, $\sigma=0.07$, dots: $\alpha_c=0.5$, $M=2$,
$\sigma=0.25$, down -- triangles: $\alpha_c=0.5$, $M=2$, $\sigma=1$, open
circles: $\alpha_c=0.3$, $M=2$, $\sigma=0.28$, up -- triangles:
$\alpha_c=0.3$, $M=6$, $\sigma=0.28$ and crosses: $\alpha_c=0$, $M=2$,
$\sigma=0.28$. The solids line refers to the theoretical
prediction~\eqref{standard}. Inset: $\upsilon$ versus $\sqrt{\rho}$, for $\rho
\in 
[0,0.5]$. This zoomed view confirms  
the correctness of the scaling dependence derived in~\eqref{small_rho}.
 } 
\label{fig3}
\end{figure}

The associated standard deviation $\upsilon$ is deduced, from a series of simulations, and shown to depend on the selected value of $\rho$. The result of the analysis 
is reported in Figure~\ref{fig3}, where the calculated value of $\upsilon$ (symbols) is plotted versus the departure density amount
$\rho$. 

For small values of the control parameter $\rho$, the standard deviation $\upsilon$ of the cluster scales proportionally to $\sqrt{\rho}$, Figure~\ref{fig2}. 
Numerics indicates that the proportionality coefficient gets smaller, as $\alpha_c$ grows. In the opposite limit, namely for large values of the density 
$\rho$, the standard deviation $\upsilon$ rapidly saturates to a asymptotic value, $\upsilon_c$. The latter is universal, meaning that it neither scales with 
$\alpha_c$, nor it does with the social temperature $\sigma$. Our best numerical estimates returns  
$\upsilon_c =0.28$. 

\begin{figure}[htbp]
\centering
\includegraphics[width=10cm]{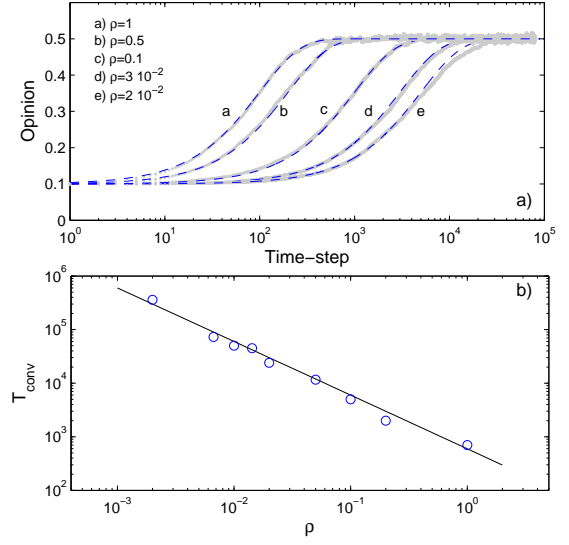}
\caption{a) The opinion as a function of time for the case where all
  the agents are initialized with the same opinion $O_o=0.1$. The solid (red)
  line represent  
the numerical data while the segmented (blue) line represents the theoretical prediction from Eq.~\eqref{mean_rOn2}. b) The transition time $T_{conv}$ as a 
function of the death density $\rho$. Symbols stand for the results of the simulations and the solid line represents the prediction~\eqref{T_{conv}}. Here $N=100$, $\alpha_c = \Delta O_c = 0 .5$, $\epsilon=0.001$ and $\sigma=0.07$.}
\label{fig4}
\end{figure}

The solid lines in Figure~\ref{fig4} represent the function :
 \begin{eqnarray}
 \upsilon^2  = \frac{M}{12 N [1- \frac{N-M}{N} (\frac{T_c-T}{T_c})^2 ] }\, ,
 \label{standard}
\end{eqnarray}
which straightforwardly follows from a simple analytical argument, developed
hereafter. In the above expression $T_c$, stands for the  
time of convergence of the opinion cluster. In~\cite{propaganda}, working
within the Deffuant's scheme~\cite{Deffuant}, i.e. in the closed
  community case without affinity, authors 
showed that the time needed 
to form a coherent assembly from a sequence of binary encounters scales as $N
\log N$, where $N$ refers to the population size. In~\cite{Bagnoli_prl}, it
was also proven that the affinity slows down the convergence rate and a
correction for $T_c$ that accounts  for the role of  
$\alpha_c$ was worked out. More precisely, it was found that :
\begin{equation}
 T_c(\alpha_c) \approx \frac{N \log(N)}{2 \log(2) \left(\alpha_{*}-\alpha_c
   \right)}\, . 
 \label{time_conv}
\end{equation}
where $\alpha_{*} \simeq 0.64$~\footnote{$\alpha_{*}$ is an effective value
  deduced via numerical fit, as discussed in~\cite{Bagnoli_prl}. A fully
  analytical derivation of $T_c$ will be presented in~\cite{carletti}.}. Before
turning to discuss the analytical derivation of
Eq.~\eqref{standard}, we wish to test its predictive adequacy with
reference to the two limiting cases outlined above. Indeed, for $\rho<<1$,  
Eq.~\eqref{standard} can be cast in the approximated form:
 \begin{eqnarray}
 \upsilon = \sqrt{\frac{T_c (\alpha_c)}{24 N}} \sqrt{\frac{M}{T}} = \gamma_t
 \sqrt{\rho}\, , 
 \label{small_rho}
\end{eqnarray}
which presents the same dependence of $\upsilon$ versus $\sqrt{\rho}$, as
observed in the numerical experiments. Moreover, 
the coefficient $\gamma_t$ is predicted to decay when increasing the cutoff in
affinity $\alpha_c$, just as expected from the numerics.  
For $\rho >> 1$,  Eq.~\eqref{standard} implies:
\begin{equation}
 \upsilon = \sqrt{\frac{1}{12}}\approx 0.28\, ,
 \label{large_rho}
\end{equation}
in excellent agreement with our numerical findings. 

To derive Eq.~\eqref{standard} let us suppose that at time  $t$ 
the death/birth process takes place and the system
experience an injection of new  
individuals. Label with $\upsilon_t$ the standard deviation of the agents
opinion distribution   
$f^t(O)$, at time $t$. It is reasonable to assume that $f^t(O)$  is centered
around $1/2$.  
After $T$ interactions between agents, when the next
perturbation will occur ($M$ agents are randomly removed from the
  community and replaced by $M$ new actors with random opinion and affinity scores) the  
distribution has been already modified, because of the underlying dynamical
mechanism specified through 
Eqs.~\eqref{opinion}. More concretely, the opinions slightly 
converge around the peak value $1/2$, an effect that certainly translates into a 
reduction of the associated standard deviation. To provide a quantitative estimate of such phenomenon, we 
recall that in the relevant $(O,t)$ plan, the convergence process fills an ideal 
{\it triangular} pattern, whose hight measures $T_c$. This topological observation enables us to put forward the following linear ansatz:
 \begin{eqnarray}
 \upsilon^{conv}_{t+T} = \frac{T_c - T}{T} \upsilon_{t}   \, ,
  \label{standard_t_plus_T_a}
\end{eqnarray}
where $\upsilon^{conv}_{t+T}$ labels the standard deviation just before the insertion of the next pool of incoming agents. 
Recalling that the newly inserted elements are uniformly distributed, and labeling with 
$G(\cdot ,\cdot)$ the opinion distribution (the two entries referring respectively to mean and the standard deviation), the updated 
variance is:
 \begin{eqnarray}
 \upsilon_{t+T}^2  &=& \frac{M}{N} \int^{1}_{0} \left(O-\frac{1}{2} \right)^2
 d O \notag \\&+& \frac{N-M}{N} \int^{1}_{0} G \left (\frac{1}{2},\upsilon^{conv}_{t+T} \right)
 \left(O-\frac{1}{2}\right)^2 dO \notag\\
 &=& \sqrt{\frac{M}{12N}} + \frac{N-M}{N} \left(\frac{T_c - T}{T}\right)^2
 \upsilon_{t}^2\, .\nonumber
  \label{standard_reborn}
\end{eqnarray}

The asymptotic stationary solution correspond to\\
$\upsilon_{t+T}=\upsilon_{t}$, a 
condition that immediately leads to Eq.~\eqref{standard} when plugged
into~\eqref{standard_reborn}. The above analysis also suggests that the final
fate of the system is not affected by the time when the  
perturbation is first applied, $t_{in}$. This conclusion is also confirmed by
direct numerical inspection:   
The asymptotic value of the standard deviation $\upsilon$ does not depend on
$t_{in}$, but solely on $\rho$. Even in the extreme condition, when the 
death/birth perturbation is switched on after
the agents have 
already collapsed to the mean opinion $0.5$,  
one observes that, after a transient, the cluster spreads and the measured
value of  $\upsilon$ agrees with the theoretical  
prediction~\eqref{standard}.

Aiming at further characterizing the system dynamics, we also studied the case
where, initially, agents share the same belief $O_0$.  
The initial distribution of opinions is therefore a Dirac delta $f^0(O)=\delta
(O-O_o)$. Such condition is   
a stationary solution, for any given $O_o$ when the
death/birth process is inactivated.   
Conversely, when the death/birth applies, the system evolves toward a final state, being
characterized by a single cluster, centered in $O=0.5$ and with  
standard deviation given by Eq.~\ref{standard}. It is also observed that the time needed by the system to complete the transition 
$T_{conv}$ depends on the value of $\rho$ and the critical affinity $\alpha_c$, see
Figure~\ref{fig4}. A simple theoretical argument enables us to quantitatively explain these findings. The initial distribution of opinions is modified 
after the first death/birth event as:
 \begin{eqnarray}
 f^1(O) =\frac{M}{N}+\frac{N-M}{N} \delta (O-O_o)\, .
  \label{f_{1}}
\end{eqnarray}
The first term refers to the freshly injected actors,  while the second stands for the remaining Delta-distributed individuals. Hence, 
the mean opinion value reads:
\begin{equation}
 \bar{O}_1 =\int^{1}_{0}{f^1(O) O dO}=\frac{M}{2N}+\frac{N-M}{N} O_o\, .
  \label{mean_O1}
\end{equation}

We can suppose that between the occurrence of two consecutive perturbations (separated by $T$ iterations),
the group average opinion  
does not significantly change. Notice that the probability of interaction of a newborn agent with another belonging to the main group is in fact proportional to M/N. Moreover several consecutive encounters of this type are necessary to induce a macroscopic change of the averaged opinion.
Under this hypothesis the next death/birth event makes the
average opinion change as:  
\begin{equation}
 \bar{O}_{2} =\frac{M}{2N}+\frac{N-M}{N} \bar{O}_1\, .
  \label{mean_O2}
\end{equation}

After $n$ death/birth iterations, the opinion
mean value reads : 
\begin{equation}
 \bar{O}_{n} =\frac{M}{2N}+\frac{N-M}{N} \bar{O}_{n-1}\, .
  \label{mean_On}
\end{equation}
From Eq.~\eqref{mean_On} one easily gets that the
asymptotic equilibrium is reached for $\bar{O}_{\infty}=0.5$, as 
 found in our 
numerical experiments; in fact the following relation is
straightforwardly obtained: 
\begin{equation}
 \bar{O}_{n} =\left(\frac{N-M}{N}\right)^{n} O_o +
 \sum^{n-1}_{l=0}{\frac{M}{2N}\left(\frac{N-M}{N}\right)^l}\, ,
  \label{mean_rOn}
\end{equation}
being $O_o$ the initial common believe. By setting
$\alpha=\frac{N-M}{N}$ 
and $\beta=\frac{M}{2N}$, the solution of Eq.~\eqref{mean_rOn} reads: 
\begin{equation}
 \bar{O}_{n} =\beta \frac{1-\alpha ^{n}}{1-\alpha}+ \alpha ^{n} O_o\, ,
  \label{mean_rOn2}
\end{equation}
whose asymptotic solution is given by $\bar{O}_{n}\rightarrow
  \bar{O}_{\infty}=0.5$. 

Expression~\eqref{mean_rOn2} reproduces quite well the dynamics of the cluster
mean, as seen in the simulations. The adequacy of~\eqref{mean_rOn2} is in fact
clearly demonstrated in Figure~\ref{fig4}a. Let us define the convergence time
$T_{conv}$  as the number of iterations needed to bring the average opinion
$\epsilon$ close to its asymptotic value $1/2$. Solving  Eq.~\eqref{mean_rOn2}
for $n$ and recalling that  $T_{conv}=nT$ yield: 
\begin{equation}
T_{conv} =  T  \log_{\alpha} \left[\frac{1}{|O_o - \frac{1}{2}|}(\epsilon - \frac{\beta}{1-\alpha }) \right]\, ,
  \label{T_{conv}}
\end{equation}
The above estimate is in excellent agreement with the numerical results reported in Figure~\ref{fig4}b.  
\section{Conclusions}
\label{sect:conclusion}
In this paper we have discussed the process of opinion making in an
open group of interacting subjects. The model postulates the coupled
dynamical evolution of both individuals' opinion and mutual affinity,
according to the rules formulated in~\cite{Bagnoli_prl}. At variance
with
respect to the toy---model~\cite{Bagnoli_prl}, the system is now open
to contact with an external reservoir of potentially interacting
candidates. Every $T$ iterations the $M$ agents are instantaneously
replaced by newborn actors, whose opinion and affinity scores
are randomly generated according to a pre--assigned (here uniform)
probability distribution. The ratio $\rho=M/T$, here termed departure
density
plays the role of a control parameter. The occurrence of a phase
transition is found which separates between two macroscopically
different
regimes: For large values of the so--called social temperature the
system collapses to a single cluster in opinion, while in the opposite
regime a fragmented phase is detected. The role of $\rho$ is
elucidated and shown to enter in the critical threshold as a linear
contribution.
Two phenomena are then addressed, with reference to the single
clustered phase. On the one side, the external perturbation, here
being hypothesized to mimic a death/birth process, induces
a spreading 
of the final cluster. The associated variance is numerically shown to
depend on the density amount
$\rho$, the functional dependence being also analytically explained.
On the other, we also show that the birth/death events
imposed at a 
constant pace can produce the progressive migration of a cluster,
initially localized around a given opinion value. A theoretical
argument is also developed to clarify this finding. As a general
comment, we should emphasize that the effect of opening up the system to
external influences changes dramatically its intrinsic dynamics
revealing peculiar, potentially interesting, features which deserves to
be further explored.

\end{document}